\documentclass[12pt,oneside,fleqn]{article}
\usepackage{epsf,graphicx}
\usepackage{amsfonts, bm}
\usepackage{amsmath}
\usepackage{amssymb}
\usepackage[latin2]{inputenc}
\usepackage[T1]{fontenc}
\usepackage{epsfig}
\usepackage{mathrsfs}
\usepackage{verbatim}

\newcommand{\e}{\varepsilon}

\newcommand{\de}{\text{d}}

\newcommand{\ket}[1]{\ensuremath{|#1\rangle}}
\newcommand{\bra}[1]{\left\langle #1 \right |}
\def\R{ {\mathbb{R}} }
\def\C{ {\mathbb{C}} }

\def\ov#1{ {\overline{#1}} }

\def\e{ {\it et al.} }
\newcommand{\al}{\alpha}
\newcommand{\h}{\mathcal{H}}
\newcommand{\z}{{\bf z}}
\newcommand{\w}{\omega}

\newtheorem{thm}{Theorem}
\newtheorem{df}{Definition}
\newtheorem{lem}{Proposition}

\newlength{\dinwidth}
\newlength{\dinmargin}
\DeclareMathAlphabet{\scr}{U}{rsfs}{m}{n}

\begin{document}

\begin{center}
{\large{\bf Statistical-mechanical description of quantum entanglement}}\\

\vspace{0.5cm}

J. K. Korbicz$^{1,2,3}$, F. Hulpke$^1$, A. Osterloh$^{1}$, and M. Lewenstein$^{1,3,4}$\\

$^1$ Institut f\"ur Theoretische Physik, 
Leibniz Universit\"at Hannover, Appelstr. 2, D-30167
Hannover, Germany

$^2$ Dept. d'Estructura i Constituents de la Mat\`eria,
Universitat de Barcelona, 647 Diagonal, 08028 Barcelona, Spain

$^3$ ICFO--Institut de Ci\`{e}ncies Fot\`{o}niques, Mediterranean Technology Park, 08860
Castelldefels (Barcelona), Spain

$^4$ ICREA-Instituci\`{o} Catalana
de Recerca i Estudis Avan\c ats, 08010
Barcelona, Spain
\end{center}

\begin{abstract}
We present a description of 
finite dimensional quantum entanglement, based on 
a study of the space of all convex decompositions of a given 
density matrix. 
On this space we construct a system of real 
polynomial equations describing separable states. We further study 
this system using methods of statistical mechanics. As an example, we apply 
finally
our techniques to Werner states of two qubits and obtain a 
sufficient criterion for separability.
\end{abstract}


\section{Introduction}
\label{statintr}
Separability is one of the central issues in quantum information theory
(see Horodecki \e \cite{RMP} for a review)
in that in a separable density matrix all correlations are of classical origin
and no real quantum information processing, as based on the 
presence of quantum entanglement of some kind, is impossible. 
The solution to the separability problem has been proved to be NP-hard~\cite{Gurvits} 
and hence every partial solution constitutes an important achievement.
Seminal corner stones in that direction have been the
Peres-Horodecki criterion~\cite{PeresHoro,Pawel},
and entanglement witnessing operators~\cite{TerhalWit,OptimWitness}.
The first method exploits the fact that positive operators 
conserve the positivity of all separable density matrices, whereas
some entangled density operators are mapped to non-positive operators.
The latter approach uses limits for expectation values of suitably chosen
witness operators to distinguish between separable and entangled states.
A systematic analysis of the so called {\em bound} entangled states has been initiated by means of
unextendible product bases (UPB)~\cite{UPBs} which in turn also paved the way towards
a formulation of the separability problem in terms of roots of complex polynomial equations~\cite{PolynEqSeparability}.
As far as we know, this route has not been pursued any further and in particular
no direct test of separability via the convex roof extension
of a pure state separability criterion has been probed so far. The main obstacles for such an approach 
have their origin in the complications involved in the minimization procedure over all decompositions
of the density matrix under consideration. A proposal in this direction however has been
presented by Osborne~\cite{Osborne04}.  
In this work we follow this route proposing a similar approach for studying the bipartite 
separability problem in finite dimensional Hilbert space
$\h=\h_A\otimes\h_B\cong\C^m \otimes \C^n$ encoding the convex roof 
minimization in a way familiar from statistical-mechanics.

The paper is organized as follows. After a formal definition of the 
separability problem and a short discussion of pure state separability
criteria in the next section, we give a geometrical view on the
space of $\rho$-ensembles and a formulation of the bipartite 
separability problem in terms of a set of nonlinear equations 
in section~\ref{rho-ens}.
A mechanical analogy of these equations is drawn in section~\ref{CostFun}
in terms of a Hamiltonian or cost function on a restricted ``phase space''
and constitutes the basis for the statistical mechanical approach presented
in section~\ref{StatMech}. After presenting a proof-of-principles 
calculation for two-qubit Werner states in section~\ref{WernerStates}
we draw our conclusions and give a short outlook of the presented formalism.

\section{The bipartite separability problem}

In order to formulate the problem, let us recall the following Definition:

\begin{df}
A state $\varrho$ of a bipartite system $AB$, described by
$\h_A\otimes\h_B$, is called separable
if there exists a convex decomposition of $\varrho$ 
composed entirely of product vectors:
\begin{equation}
\varrho=\sum_{i=1}^{N}p_i \,|x_i\rangle\langle x_i|
\otimes|y_i\rangle\langle y_i|, \quad \ket{x_i}\in\h_A, \ket{y_i}\in\h_B. 
\end{equation}
\end{df}

A natural problem arises, known as the separability problem: 
{\it Given a state $\varrho$, decide if it is
separable or not.} This problem has been proven to be NP-hard
(Gurvits~\cite{Gurvits}) and (a part of) its difficulty
lies in the fact that a
convex decomposition of a given mixed state $\varrho$ into pure states:
\begin{equation}\label{decomp}
\varrho=\sum_{i=1}^{N}p_i \,|\Psi_i\rangle\langle \Psi_i|
\end{equation}
is highly non-unique (see e.g. Bengtsson and $\dot{\text{Z}}$yczkowski~\cite{Bengtsson}). 
Thus, the following Definition makes sense:

\begin{df}\label{roens}
Unordered collection $\{p_i, \ket{\Psi_i}\}$, $i=1\dots N$ of probabilities 
and vectors satisfying (\ref{decomp}) is called a $\varrho$-ensemble
of length $N$.
\end{df}

In this work we develop the following approach to the 
separability problem:
we propose to search the space of all $\varrho$-ensembles of a given state 
$\varrho$ for product $\varrho$-ensembles
($\varrho$-ensembles containing only product vectors), 
by applying one of the existing necessary and sufficient 
entanglement tests to each member of the ensemble.
We want the test which has the simplest functional 
form---a polynomial. Such a test is provided by
the square of generalized concurrence 
(see e.g. Rungta \e~\cite{Rungta},
Mintert \e~\cite{Mintert}, Hulpke~\cite{Florek}):

\begin{lem}\label{warprod}
For any vector $\ket\psi\in\h_A\otimes \h_B$ one has that:
\begin{equation}\label{test}
c^2(\psi):=||\psi||^4 - \rm{tr}_{\h_A}(\rm{tr}_{\h_B} |\psi\rangle\langle \psi|)^2\ge 0
\end{equation} 
and the equality holds if and only if $\ket\psi$ is product.
\end{lem}

This leads to a set of real polynomial equations describing
separable states.
The resulting system is  
very complicated due to the fourth order of some equations 
and a large number of variables. Our idea is to study it 
using methods of classical statistical mechanics.
The motivation is that such methods have proven to be very efficient 
not only within classical mechanics, but also  
in many other, distantly related areas (for an application
to fundamental combinatorial problems see e.g. Kubasiak \e
\cite{Kubasiak} and references therein).
Hence, we first develop a mechanical analogy for our system.
Then we define a suitable cost function, or ``energy'', 
introduce a canonical ensemble, and study the resulting partition function. 

\section{The space of $\varrho$-ensembles and separability}
\label{rho-ens}
Let us begin with describing the space of all $\varrho$-ensembles of a given
state $\varrho$.
For convenience we pass from normalized $\varrho$-ensemble vectors $\ket{\Psi_i}$ 
to subnormalized ones: $\ket{\psi_i}:=\sqrt{p_i}\ket{\Psi_i}$, such that 
$\varrho=\sum_{i=1}^{N}|\psi_i\rangle\langle \psi_i|$. Let us fix 
an eigenensemble $\{\ket{e_\alpha}\}$ of $\varrho$, where 
all the vectors $\ket{e_\al}$ correspond to non-zero eigenvalues 
$\lambda_\al$ of $\varrho$, $\alpha=1\dots r$, and 
$r:=\text{rank}(\varrho)$ is the rank of $\varrho$. Then, 
all $\varrho$-ensembles are characterized by the well known
Theorem by Schr\"odinger~\cite{Schrod} 
(see also~\cite{Hugh,Kirkpatrick})

\begin{thm}\label{HJW}
Any  
$\varrho$-ensemble 
$\{\ket{\psi_i}\}$ of length $N\ge r$ can be 
obtained from a subnormalized eigenensemble $\{\ket{e_\alpha}\}$ 
such that $\rho=\sum_\alpha \ket{e_\alpha}\bra{e_\alpha}$ through the following linear transformation:
\begin{equation}\label{ens}
\ket{\psi_i}:=\sum_{\alpha=1}^r z_{i\alpha} \ket{e_\alpha},
\end{equation}
where the matrix $z_{i\alpha}\in \mathbb{C}$ is an $N\times r$ block of a unitary $N\times N$ matrix, and hence satisfies
\begin{equation}\label{stiefel}
\sum_{i=1}^{N} \ov{z_{i\alpha}}z_{i\beta}=\delta_{\alpha\beta}.
\end{equation}
\end{thm}

%

Theorem~\ref{HJW} gives us the characterization of all possible
$\varrho$-ensembles in terms of $N\times r$ matrices $z$, satisfying
the condition (\ref{stiefel}). Geometrically, this condition
defines the so called {\em Stiefel manifold }
\begin{equation}\label{st}
V_{N,r}:=U(N)/U(N-r).
\end{equation}
It forms a principal fiber bundle over the Grassmann manifold $G_{N,r}$ 
(the set of $r$-dimensional subspaces of $\C^N$) with a fiber 
diffeomorphic to $U(r)$
(we refer to Kobayashi and Nomizu Vol.~1~\cite{Kobayashi1} for the 
definition and basic properties of fiber bundles and to 
Spivak Vol. 5~\cite{Spivak} for more
information on the Stiefel and Grassmann manifolds). 

However, note that there is some
additional symmetry: from Eq.~(\ref{decomp}) we see that the order of 
vectors in a $\varrho$-ensemble does not matter, and thus two 
$N \times r$ matrices $z$, $z'$ satisfying Eq.~(\ref{stiefel}) and
differing only by a permutation of their rows define the 
same $\varrho$-ensemble. 
To fix this freedom, observe that a $z$-matrix satisfying 
Eq.~(\ref{stiefel}) has necessarily rank $r$, and hence we may consider 
only those matrices $z$, for which the first $r$ rows 
are linearly independent. 
The set of such $z$'s constitutes a simply connected open subset of 
$V_{N,r}$ (which is nevertheless dense in $V_{N,r}$) and over such a  
neighborhood the bundle  
$V_{N,r}\xrightarrow []{U(r)}G_{N,r}$ is trivial by construction.
This allows us to formally write down an explicit solution
of the constraints (\ref{stiefel})
\begin{equation}\label{Srozw}
z=GS\left( \begin{array}{c}{\bf 1}_r \\ {\bf v} \end{array}\right)\cdot 
U, 
\end{equation}
where $U \in U(r)$, ${\bf 1}_r$ is the $r \times r$ unit matrix, 
${\bf v}$ is an arbitrary, 
complex $(N-r) \times r$ matrix, and $GS$ denotes the 
Gram-Schmidt orthonormalization~\cite{GS} applied to the columns. 
There are no more symmetries, since we have
defined in Definition~\ref{roens} $\varrho$-ensembles using vectors 
$\ket{\psi_i}$ rather than more physical projectors 
$\ket{\psi_i}\langle\psi_i|$, as the latter are harder to work with. 
In case of $\varrho$-ensembles defined through projectors, there would 
be an additional symmetry of multiplying each row of $z$ by a (different)
phase. Comparing Eq.~(\ref{Srozw}) and Eq.~(\ref{ens}), one sees that an
arbitrary $\varrho$-ensemble of length $N$ is obtained from the fixed
eigenensemble by i) applying a unitary rotation to $\ket{e_\alpha}$'s
and ii) subsequent increasing of the length of the
ensemble along the Grassmannian $G_{N,r}$.

So far we have characterized $\varrho$-ensembles of a fixed 
length $N$. It seems that in the search for product ensemble
we would have to consider all possible lengths
$N\ge r$. However,  from Caratheodory's Theorem 
(see e.g. Kelly and Weiss~\cite{Car}) we know that 
a separable state can 
be decomposed into at most $N=m^2n^2$ linear independent 
(in $\R^{m^2n^2-1}$) product states. 
Hence, 
it is enough to consider only $\varrho$-ensembles of the length
$N=m^2n^2$.
(there is a natural inclusion of space of shorter ensembles in the space of longer ones).

Let us now examine the entanglement test given by 
Proposition~\ref{warprod}. First, we quote
some well known facts regarding the geometry of
pure product states 
(see e.g. Bengtsson and $\dot{\text{Z}}$yczkowski~\cite{Bengtsson}).  
Note that the
polynomial $c^2(\psi)$, defined in Eq.~(\ref{test}),
is in fact a sum of modulus squared of quadratic, 
complex-analytical polynomials in $\ket\psi$:
\begin{equation} \label{c2}
c^2(\psi)=\frac{1}{2}\sum_{a,b=1}^{d_1,d_2}
\big|\langle \zeta_a^{AA'}\otimes \widetilde \zeta_b^{BB'}|\psi^{AB}\otimes\psi^{A'B'}\rangle\big|^2,
\end{equation}
where $\{\ket{\zeta_a^{AA'}}\}_{a=1,\dots,d_1}$, 
$\{\ket{\widetilde \zeta_b^{BB'}}\}_{b=1,\dots,d_2}$ 
are orthonormal bases of the skew-symmetric 
spaces $\h_A\wedge\h_{A'}\cong\C^m \wedge \C^m$ 
and $\h_B\wedge\h_{B'}\cong\C^n \wedge \C^n$, respectively.
Thus, $c^2(\psi)=0$, and hence $\ket\psi$ is product, if and only if
\begin{equation}\label{segre}
\langle \zeta_a\otimes \widetilde \zeta_b|
\psi\otimes\psi\rangle=0\quad \text{for all} \quad a,b\; .
\end{equation} 
It is worth noticing that this is just the condition for the matrix of components of $\ket\psi$ to have rank one. 
Geometrically, the system of homogeneous equations (\ref{segre}),
or equivalently the single equation $c^2(\psi)=0$,  
describes the image of the so called {\em Segre embedding}
$\C P^m \times \C P^n\hookrightarrow \C P^{mn}$ given by
$([x],[y])\mapsto [x\otimes y]$.
As we can see from Eqs.~(\ref{segre}),
this image, i.e. the set of product vectors, is a complex-analytical 
manifold---as an intersection of complex quadrics---in contrast 
to the Stiefel manifolds $V_{N,r}$, which are real.

Since for all $i=1,\dots N$ polynomials $c^2(\psi_i)$ are non-negative and equal to zero
if and only if $\ket{\psi_i}$ is product, we can sum them up
for a given $\varrho$-ensemble, and thus obtain a
collective separability test for the whole $\varrho$-ensemble,
given by a single polynomial function. Combining this with the 
parametrization (\ref{ens}) and the constraint (\ref{stiefel}), 
we obtain the following description of separable states: 

\begin{lem}\label{p2}
A states $\varrho$ of rank $r$ on $\C^m\otimes\C^n$ 
is separable if and only if
the following system of 
equations possesses a solution 
\begin{eqnarray}
& &E_\varrho(z):=\sum_{i=1}^{m^2n^2}c^2(\psi_i)
=\sum_{i=1}^{m^2n^2} \sum_{\alpha, \dots ,\nu=1}^r
\ov{z_{i\alpha}}
\:\ov{z_{i\beta}}E^\varrho_{\alpha\beta\mu\nu}z_{i\mu}z_{i\nu}=0,\label{H}\\
& & \mathcal{C}_{\al\beta}(z):= \sum_{i=1}^{m^2n^2} \ov{z_{i\alpha}}z_{i\beta}-
\delta_{\alpha\beta}=0,\label{wiaz}
\end{eqnarray}
where
\begin{equation}\label{CostOp}
E^\varrho_{\alpha\beta\mu\nu}:=\frac{1}{4}\langle e_{\alpha}\otimes e_{\beta}|
\Pi_m \otimes \Pi_n \,e_{\mu}\otimes e_{\nu}\rangle
\end{equation}
and $\Pi_m,\Pi_n$ are the projectors from $\C^m\otimes \C^m$, 
$\C^n\otimes \C^n$ onto the skew-symmetric 
subspaces $\C^m \wedge \C^m$, $\C^n \wedge \C^n$ respectively. 
\end{lem}

We note that the pure state entanglement measure we use is the square of the  
generalized concurrence $c(\varrho)$ (cf. Rungta \e~\cite{Rungta},
Mintert \e~\cite{Mintert}),
however, instead of the convex roof construction
$c(\varrho):=\text{inf}\sum_i p_ic(\Psi_i)=\text{inf}\sum_i c(\psi_i)$
(where $\ket{\Psi_i}$ are normalized vectors),
we analyze
\begin{equation}\label{p^2}
E_\varrho(z)=\sum_i p_i^2c^2(\Psi_i). 
\end{equation}
as a "quantifier" of entanglement. We remind the reader that no caveats are introduced by this,
since we are only interested in the detection of zero entanglement rather than in the full construction
of an entanglement monotone.

Note that the Eqs.~(\ref{H}), (\ref{wiaz}) are invariant with respect to
local unitary transformations, since when 
$\varrho$ is separable also $U_A\otimes U_B\varrho U_A^\dagger\otimes U_B^\dagger$ is, 
for arbitrary 
$U_A\in U(m)$, $U_B\in U(n)$. 
The latter transformation can be viewed either as
a local change of basis (passive view) or as an active rotation (active view).
Indeed, from Eq.~(\ref{test}) one immediately sees that 
$c^2(U_A\otimes U_B \psi)=c^2(\psi)$. Thus, the function
$E_\varrho$ and all quantities derived from it
are constant on the whole unitary class of $\varrho$, i.e. on
$[\varrho]:=\{U_A\otimes U_B\varrho U_A^\dagger\otimes U_B^\dagger\, ; \,
U_A\in U(m), U_B\in U(n)\}$. 
In what follows, we refer with $\varrho$ to its local 
unitary class $[\varrho]$.

We give a brief comparison to a previous analysis carried out by Wu \e in Ref.~\cite{Wu},
also leading to a different set of polynomial equations. These 
authors have used a higher order polynomial test for separability:
let $\sigma_A := \text{tr}_{\h_B} |\psi \rangle \langle \psi|$, then 
$\ket\psi$ is product if and only if $\text{det}(\sigma_A - {\bf 1})=0$. 
The relation to Eq.~(\ref{test}) is established by observing 
that $\text{det}(\sigma_A - {\bf 1})=\sum_{k=0}^{m}(-1)^k c_k(\sigma_A)$, 
where $c_k$'s form a basis of $U(m)$-invariant polynomials 
(see e.g. Ref.~\cite{Kobayashi}). Particularly, 
$2c_2(\sigma_A)=\big( \text{tr}\sigma_A \big)^2-\text{tr}\sigma_A^2$, 
which is precisely the generalized concurrence squared (cf. Eq.~(\ref{test})). 
For testing separability, it is sufficient to consider only $c_2$.

\section{Mechanical analogy} \label{CostFun}
Equations (\ref{H}) and (\ref{wiaz}) 
form a system of real (after taking real and imaginary parts)
polynomial equations. Let us denote by $\mathcal{V}_\varrho$ the set of its solutions
for a given $\varrho$.
Then the separability problem is equivalent to the question whether
$\mathcal{V}_\varrho$ is empty or not. In principle there
is a general solution to such problem, provided by the so called
{\em Real Nullstellensatz} (see e.g. Bochnak \e~\cite{Bochnak}). 
It says that $\mathcal{V}_\varrho=\emptyset$ if and
only if the ideal generated by the polynomials
$E_\varrho$, $\{\text{Re}\mathcal{C}_{\alpha\beta},\text{Im}\mathcal{C}_{\alpha\beta}\}$,
and by all (real) sum-of-squares (SOS) polynomials\footnote{
Interestingly, SOS polynomials also appear in a 
solution to the classicality problem of states of a single mechanical
system---they are enough to detect a very broad family of states
through generalized squeezing conditions (Korbicz \e~\cite{hp17}).}
contains the constant $-1$. Equivalently, $\mathcal{V}_\varrho=\emptyset$
if and only if there exist a SOS 
polynomial $s=\sum_n(w_n)^2$, a real polynomial $t$, and (complex) polynomials 
$u_{\al\beta}$ such that: 
\begin{equation}\label{Gcertif}
-1=s(z)+E_\varrho(z) t(z)+ 
\sum_{\al,\beta}\text{Re}\big[\mathcal{C}_{\alpha\beta}(z)\,\ov{u_{\al\beta}(z)}\big].
\end{equation} 
However, finding such
a certificate is computationally very difficult and inefficient, 
due to the fact that the degrees of polynomials 
$s$, $t$, and $u_{\al\beta}$ are a priori unbounded 
(see also Refs.~\cite{Osborne04,Badziag}).

Here we develop a different approach based on a statistical analysis of a 
classical-mechanical analogy. Namely, we treat $z_{i\alpha}$ 
as a collection of complex row vectors 
$\z_i\in\mathbb{C}^r\cong\R^{2r},\,i=1\dots N$ and treat each row $\z_i$ as 
a complex phase-space coordinate of a fictitious particle 
moving in $r$-dimensional space. Then the whole matrix $z_{i\alpha}$ becomes a 
phase-space coordinate of a system of $N$ such particles in
their composite phase-space $\Gamma:=\R^{2r}\times\cdots\times\R^{2r}\cong\C^{Nr}$. 
Now, let $E_{\varrho}(\z_1,\dots,\z_N)$ and 
$\mathcal{C}_{\alpha\beta}(\z_1,\dots,\z_N)$ be defined by Eqs.~(\ref{H}) and 
(\ref{wiaz}). We emphasize that $E_\varrho$ depends on the separability class of the analyzed 
state $\varrho$ through the fixed eigenensemble $\{\ket{e_{\alpha}}\}$. 
From the property (\ref{test}) it follows that
\begin{equation}
E_{\varrho}(\z_1,\dots,\z_N)\ge 0 \quad \text{for any}\quad (\z_1,\dots,\z_N)\in\C^{Nr}.
\end{equation}  
We will think of $E_\varrho$ as a cost function or Hamiltonian
(it is extensive in the number of fictitious particles $N$), 
of our fictitious mechanical system. Then, we can treat 
$\mathcal{C}_{\alpha\beta}$ as at the primary 
constraints imposed on the 
a priori independent phase-space 
coordinates $(\z_1,\dots,\z_N)$. 
We note that even if the mechanical system corresponds to free particles (if $E_{\varrho}$ was diagonal)
the resulting model is nevertheless interacting due to the forces of inertia induced by the non-linear constraints.

The corner stones of the mechanical interpretation of the separability problem (\ref{H}), (\ref{wiaz}) can be summarized
as follows: 
the $\varrho$-ensembles of
density matrices with a fixed rank $r$ form the Stiefel manifold 
$V_{N,r}$, which we can be viewed at as a constraint surface in the phase-space
$\Gamma$. Each state $\varrho$ defines the non-negative 
cost operator $E^\varrho_{\alpha\beta\mu\nu}$ \eqref{CostOp} 
which uniquely defines the cost function $E_\varrho$ on $\Gamma$, which probes the separability of the ensembles.
The cost function $E_\varrho$ assumes the value zero on the constraint surface $V_{N,r}$ (which is then its global
minimum) if and only if $\varrho$ is separable.

\section{Statistical-mechanical approach}\label{StatMech}
Although in principle one could tempt to solve the 
constraints explicitly by Eq.~(\ref{Srozw}), the
resulting parametrization of the constrained manifold 
is rather hard to work with due to the iterative nature of the 
Gram-Schmidt orthonormalization.
We circumvent the complications with an explicit incorporation of the constraints by 
using a standard method of implicit treatment of constrained systems due to
Dirac~\cite{Dirac}. It is based on the introduction of Lagrange multipliers. 
To this end we define the full Hamiltonian of the systems as 
\begin{eqnarray}
H_{full}(\z_1\dots \z_N):=E_{\varrho}
+\sum_{\alpha,\beta}\w_{\alpha\beta}\mathcal{C}_{\alpha\beta},\label{HT}
\end{eqnarray}
where $\w_{\alpha\beta}$ are the Lagrange multipliers. 
Note that the constraints written in 
the matrix  $\mathcal{C}_{\alpha\beta}$ are not all independent: it is in fact a hermitian matrix
and we need to employ one Lagrange multiplier for each independent constraint only.
On the other hand we have considerable freedom for choosing the spurious Lagrange multipliers in the Lagrange matrix $\w$.
We choose $\w$ to be hermitian. Then, $H_{full}$ is hermitian and has only real eigenvalues. 
Moreover, in order to take into account all independent constraints, we require 
that $\text{det}\,\w\ne0$.
The constraints $\mathcal{C}_{\al\beta}\equiv 0$ are then realized on average by setting 
to zero the variation of $H_{full}$ with respect to $\w_{\alpha\beta}$\,: 
$\partial H_{full}/\partial \w_{\alpha\beta}=0$. 

The number of fictitious particles $N$ will in general be 
notably large---in dimension $2\otimes 4$ for example 
we have $N\ge 64$. 
Thus, the direct analytical study of our fictitious 
mechanical system seems rather hopeless and 
we proceed further using methods of statistical mechanics
and numerical simulations. 
The most natural framework would be microcanonical ensemble, 
however it is also difficult to work with. Hence, we will 
introduce a canonical ensemble, keeping in mind that this is just 
a technical tool, so, for example, the inverse temperature $\beta$
plays only a  
role of a parameter here, without any physical meaning.

We proceed to define the canonical partition function $Z$ for 
our system. The most natural definition is perhaps the following
\begin{eqnarray}
& & Z(\beta;\varrho)=\int \prod_{i,\mu}\de ^2 z_{i\mu} 
\prod_{\al\leq\beta}\delta\big[\mathcal{C}_{\al\beta}(\z_1\dots \z_N)\big]\text{e}^{-\beta 
E_{\varrho}}\nonumber\nonumber\\
& & =\int \prod_{i,\mu}\de ^2 z_{i\mu} 
\prod_{\al\leq\beta}\delta\bigg[\sum_{i=1}^{N} \ov{z_{i\alpha}}z_{i\beta}-
\delta_{\alpha\beta}\bigg]\text{exp}
\bigg\{\!\!-\beta\sum_{i=1}^{N} \sum_{\alpha, \dots ,\nu=1}^r
\ov{z_{i\alpha}}
\:\ov{z_{i\beta}}E^\varrho_{\alpha\beta\mu\nu}z_{i\mu}z_{i\nu}\bigg\},\label{Zinna}
\end{eqnarray}
where the integration is explicitly restricted to the constraint
surface $V_{N,r}$ (cf. Eq.~(\ref{st})) given by $\mathcal{C}_{\al\beta}\equiv 0$.
The intuition behind such an approach is the following.
We can formally introduce constraint ``state density'' function
\begin{equation}\label{statedens}
\rho(\epsilon):=\int \prod_{i,\mu}\de ^2 z_{i\mu} 
\prod_{\al\leq\beta}\delta\big[\mathcal{C}_{\al\beta}(\z_1\dots \z_N)\big]
\delta\big(\epsilon-E_\varrho(z)\big).
\end{equation}
Since $E_\varrho(z)\ge 0$, $\rho(\epsilon)$ is non-zero only
for $\epsilon\ge 0$. Then:
\begin{equation}
Z(\beta;\varrho)=\int_0^\infty \de \epsilon \rho(\epsilon)
\text{e}^{-\beta \epsilon}.
\end{equation}
Let us assume that the state in question is entangled.
Then $E_\varrho(z)$ is strictly positive, so there exists
a constant $a$ such that $E_\varrho(z)\ge a>0$. The average
``energy'' is then separated from zero:
\begin{eqnarray}\label{saturation}
\langle\langle E_\varrho\rangle\rangle :=\frac{1}{Z(\beta;\varrho)}
\int_0^\infty \de \epsilon \rho(\epsilon)\,\epsilon
\text{e}^{-\beta \epsilon}
\ge\frac{1}{Z(\beta;\varrho)}
\int_a^\infty \de \epsilon \rho(\epsilon)a
\text{e}^{-\beta \epsilon}=a.
\end{eqnarray}
Now let $\varrho$ be separable. Then, by Proposition \ref{p2}
$E_\varrho(z)$ has zeros on the constraint surface
$\mathcal{C}_{\al\beta}=0$ with each zero corresponding to 
a separable $\varrho$-ensemble. Since such ensembles are ``rare'',
we expect that the state density $\rho(\epsilon)\to 0$ with
$\epsilon\to 0$. Let us assume for a moment that the leading term in 
the actual
dependence of $\rho(\epsilon)$ was given by a power law
\begin{equation}\label{ansatz}
\rho(\epsilon)=A\epsilon^\delta, \quad A,\delta>0.
\end{equation}
Then we obtain the well established result
$Z(\beta;\varrho)=\frac{A}{\beta^{\delta+1}}\Gamma(\delta+1)$
and 
\begin{equation}
\langle\langle E_\varrho\rangle\rangle 
=\frac{\Gamma(\delta+2)}{\Gamma(\delta+1)}\,\frac{1}{\beta}=\frac{\delta+1}{\beta}\label{1/b}.
\end{equation}
Thus, we put forward the following conjecture:

{\bf Conjecture.} {\it For the "state density" function $\rho(\epsilon)$, 
defined in Eq. (\ref{statedens}), it holds:
i) the mean energy (defined in Eq. (\ref{saturation})) 
$\langle\langle E_\varrho\rangle\rangle=a>0$ if and only if $\varrho$ is entangled;
ii) the mean energy $\langle\langle E_\varrho\rangle\rangle$ 
scales as $1/\beta$ if and only if $\varrho$ is separable.} 

We anticipate that indeed we observe such a behavior 
in a simple case of $2\otimes 2$ Werner states~\cite{Werner}.
Note that in general the exponent $\delta$ will depend on the state
$\delta=\delta(\varrho)$.

The partition function defined in Eq.~(\ref{Zinna}) is difficult to work with
analytically (however one can still investigate it numerically, e.g. using
Monte Carlo methods),
so we use a different object---the partition function for the full
Hamiltonian  (\ref{HT}).
We first rescale the variables: 
$z_{i\alpha}\mapsto z_{i\alpha}/\sqrt{N}$ and then define: 
\begin{eqnarray}
Z(\beta,\w;\varrho):=\int \prod_{i,\mu}\de ^2 z_{i\mu} \,
\text{exp}\Big[-\frac{\beta}{N^2} \Big(E_{\varrho}(\z_1\dots \z_N)
+N\sum_{i}\langle \z_i|\w \z_i\rangle-N^2\text{tr}\w\Big)\Big],
\label{Z_0}
\end{eqnarray}
where $\langle\,\cdot\,|\,\cdot\,\rangle$ denotes the standard scalar 
product in $\mathbb{C}^r$. Performing further rescaling:
\begin{equation}
\beta=N^2\tilde\beta \, , \quad \w=\frac{N}{\beta}\tilde\w \,,\label{skala}
\end{equation}
$Z(\beta,\w;\varrho)$ becomes (after dropping the tildes):
\begin{eqnarray}
& &Z(\beta,\w;\varrho)=\int \prod_{i,\mu}\de ^2 z_{i\mu} 
\,\text{exp}\Big[-{\beta}E_{\varrho}(\z_1\dots \z_N)
 -\sum_{i}
\langle \z_i|\w \z_i\rangle+N\text{tr}\w\Big]\nonumber\\
& &=\int \prod_{i,\mu}\de ^2 z_{i\mu} 
\,\text{exp}\Big[-{\beta}\sum_i \sum_{\alpha, \dots ,\nu}
\ov{z_{i\alpha}}
\:\ov{z_{i\beta}}E^\varrho_{\alpha\beta\mu\nu}z_{i\mu}z_{i\nu}
-\sum_{i}\langle \z_i|\w \z_i\rangle+N\text{tr}\w
\Big].\label{Z_stare}
\end{eqnarray}
Now we are able to reproduce the (rescaled) 
constraints (\ref{wiaz}) only on average: 
\begin{equation}
\frac{\partial}{\partial \w_{\alpha\beta}}\,
\text{log}Z(\beta,\w;\varrho)=
\langle\langle N\delta_{\alpha\beta}-
\sum_i \ov{z_{i\alpha}}z_{i\beta}\rangle\rangle
\label{constr_stare_stare} 
\end{equation}
where the average $\langle\langle \:\cdot\: \rangle\rangle$ is taken 
with respect to the probability density defined through 
Eq.~(\ref{Z_stare}): 
\begin{eqnarray}
P_{\varrho}(\z_1\dots \z_N;\beta,\w)&:=&\frac{1}{Z(\beta,\w;\varrho)}
\,\text{exp}\Big[-{\beta}E_{\varrho}(\z_1\dots \z_N)
-\sum_{i}\langle \z_i|\w \z_i\rangle+N\text{tr}\w\Big].
\end{eqnarray}
Thus, requiring that $\partial/\partial \w_{\alpha\beta}\,
\text{log}Z(\beta,\w;\varrho)=0$ amounts to:
\begin{equation}\label{constr_stare}
N\delta_{\alpha\beta}=
\langle\langle\sum_i \ov{z_{i\alpha}}z_{i\beta}\rangle\rangle. 
\end{equation}
Following the standard treatment of 
constrained systems, the equations (\ref{constr_stare}) 
are treated as conditions imposed on a priori 
arbitrary (apart form being hermitian and non-singular) matrix 
of Lagrange multipliers $\omega$.
We note that the above approach based on $H_{full}$ is nothing else but
a (formal) evaluation of the integral (\ref{Zinna}) through
the saddle point method with $N\to \infty$.

A significant simplification of the partition function 
(\ref{Z_stare}) comes from the form of our Hamiltonian $E_\varrho$---from 
Eq.~(\ref{H}) it follows that $E_\varrho(\z_1\dots \z_N)=
\sum_i E_{\!1\varrho}(\z_i)$, where $E_{\!1\varrho}$ is just 
the function $E_{\varrho}$ with $N=1$.  
The situation is more subtle with the
constraints (\ref{constr_stare}).
For the purpose of this work we will assume
that the contribution to the sum from each fictitious particle 
is equal,i.e. $\langle\langle\ov{z_{i\alpha}}
z_{i\beta}\rangle\rangle=\delta_{\alpha\beta}$
for every $i$. In general, such ``equipartition'' of course
does not have to hold and it is an additional restriction
on the Lagrange multipliers.
By such an assumption we however achieve a factorization
of the partition function:
\begin{equation}
Z(\beta,\w;\varrho)=[Z_1(\beta,\w;\varrho)]^N,
\end{equation} 
where $Z_1$ is a one-particle partition function:
\begin{eqnarray}
& &Z_1(\beta,\w;\varrho):=\int \prod_{\mu=1}^{r}\de ^2 z_{\mu} 
\,\text{exp}\Big[-{\beta}E_{\! 1\varrho}(\z)
-\langle \z|\w \z\rangle + \text{tr}\w \Big]\nonumber\\
& &=\int \prod_{\mu=1}^{r}\de ^2 z_{\mu} 
\,\text{exp}\Big[-{\beta}\sum_{\alpha, \dots ,\nu}
\ov{z_{\alpha}}
\:\ov{z_{\beta}}E^\varrho_{\alpha\beta\mu\nu}z_{\mu}z_{\nu}
-\langle \z|\w \z\rangle + \text{tr}\w\Big].\label{Z}
\end{eqnarray}

From now on we will consider $Z_1$ only. 
The constraint 
equations (\ref{constr_stare}) are then replaced by 
a one-particle version:
\begin{equation}
\frac{\partial}{\partial \w_{\alpha\beta}}\,
\text{log}Z_1(\beta,\w;\varrho)=\delta_{\alpha\beta}-
\langle\langle \ov{z_{\alpha}}
z_{\beta} \rangle\rangle =0\ , \label{constr} 
\end{equation}
in accordance with our extra assumption made above.
The average in Eq. (\ref{constr}) 
is taken with respect to the 
probability distribution:
\begin{equation}
P_{1\varrho}({\bf z};\beta,\w):=\frac{1}{Z_1(\beta,\w;\varrho)}
\,\text{exp}\Big[-{\beta}E_{\! 1\varrho}(\z)
-\langle \z|\w \z\rangle+ \text{tr}\w\Big] . \label{SP}
\end{equation}
In particular, Eq.~(\ref{constr}) implies that
$\langle\langle |z_{\alpha}|^2\rangle\rangle =1$.

To understand the meaning of Eq.~(\ref{constr}), 
let us assume that $\w=\w_0(\beta;\varrho)$ is 
its solution. 
Then Eq.~(\ref{constr}) implies that a family of 
vectors $\{\ket{\psi(\z)}:=\sum_{\alpha} z_{\alpha} \ket{e_{\alpha}}\, ;\, 
\z\in \mathbb{C}^r\}$ forms a continuous $\varrho$-ensemble 
with respect to the probability distribution (\ref{SP}), i.e.: 
\begin{equation}\label{cont_ens}
\int \! \de ^{2r} \z P_{1\varrho}(\z;\beta,\w_0)\,
| \psi(\z)\rangle \! \langle \psi(\z)| = \varrho
\end{equation}
irrespectively of $\beta$. 
Since $E_{\! 1\varrho}(\z)=c^2\big(\psi(\z)\big)$ 
(cf. Eqs.~(\ref{test}) and (\ref{H}))
is the concurrence squared of each $\ket{\psi(\z)}$, 
the average ``energy'' is just
the ensemble average of the concurrence squared:
\begin{eqnarray}
\langle\langle E_{\! 1\varrho} \rangle\rangle_0(\beta):=
\int \de^{2r} \z \, P_{1\varrho}\big[\z;\beta,\w_0(\beta;\varrho)\big]\,
E_{\! 1\varrho}(\z)
=-\frac{\partial}{\partial \beta}\,
\text{log}Z_1\Big|_{\w=\w_0(\beta;\varrho)}.\label{avH} 
\end{eqnarray}

Due to the property (\ref{c2}) one can formally simplify 
the integral (\ref{Z}) using the Hubbard-Stratonovitch trick. 
Indeed, Eq.~(\ref{Z}) can be rewritten as:
\begin{eqnarray}
Z_1(\beta,\w;\varrho)=
\int \prod_{\mu=1}^{r}\de ^2 z_{\mu}
\text{exp}\bigg\{-{\beta}\sum_{a,b=1}^{d_1,d_1}
\bigg|\sum_{\alpha, \beta}h^{ab}_{\alpha\beta}(\varrho)z_\alpha z_\beta\bigg|^2 
-\langle \z|\w \z\rangle+\text{tr}\w\bigg\},
\end{eqnarray}
where: 
\begin{equation}\label{hab}
h^{ab}_{\alpha\beta}(\varrho):=\langle \zeta_a\otimes 
\widetilde \zeta_b|e_\alpha\otimes e_\beta\rangle
\end{equation} 
and we have rescaled $\beta$ by $1/4$.
Next, we use the Hubbard-Stratonovitch substitution:
\begin{equation}
\text{exp}(-{\beta}|y|^2)=\int \frac{\de ^2 s}
{\pi\beta}\,\text{exp}\left(\!\!-\frac{|s|^2}{\beta}
+\text{i}\:\ov s y+\text{i}s\ov y\right). 
\end{equation}
to obtain (after a formal interchange of the integrations):
\begin{eqnarray}
& &Z_1(\beta,\w;\varrho)=\int\prod_{a,b=1}^{d_1,d_2}\frac{\de^2 s_{ab}}{\pi\beta}\,
\text{exp}\left(-\frac{1}{\beta}\sum_{a,b}|s_{ab}|^2+\text{tr}\w\right)
\int \frac{1}{2^r}\prod_{\mu=1}^{r}\de z_{\mu}\de \ov{z_{\mu}}\nonumber\\
& &\times\text{exp}\bigg\{\sum_{\alpha,\beta}\Big[- \ov{z_{\alpha}}\w_{\alpha\beta}z_\beta
+\text{i}\sum_{a,b}\ov{s_{ab}}\,h^{ab}_{\alpha\beta}(\varrho)z_\alpha z_\beta
+\text{i}\sum_{a,b}s_{ab}\,\ov{h^{ab}_{\alpha\beta}(\varrho)}\ov{z_\alpha}\, \ov{z_\beta}
\Big]\bigg\}.
\end{eqnarray}
The above integral is finite if and only if
$\omega> 0$ (as we said earlier we assume $\omega$ to be non-singular
in order not to loose any of the constraints, 
hence the strong inequality here). This puts no restriction on the 
amount of independent parameters in $\omega$ and from now on we 
will assume this condition to hold. Performing the Gaussian integration 
in the $2r$ variables ${\bf z},{\bf \ov z}$ 
finally yields:
\begin{eqnarray}
Z_1(\beta,\w;\varrho)=\pi^r \int\prod_{a,b=1}^{d_1,d_2}\frac{\de^2 s_{ab}}{\pi\beta}
\text{exp}\Big(-\frac{1}{\beta}\sum_{a,b}|s_{ab}|^2+\text{tr}\w\Big)
\frac{1}{\sqrt{\text{det}M_\varrho({\bf s},\w)}},\label{poHS}
\end{eqnarray}
where $2r\times 2r$ matrix $M_\varrho({\bf s},\w)$ is defined as follows:
\begin{equation}\label{macierz}
M_\varrho({\bf s},\w):=\left[\begin{array}{cc}
\omega & -2\text{i}\sum_{a,b}s_{ab}\ov{{\bf h}^{ab}(\varrho)}\\
-2\text{i}\sum_{a,b}\ov{s_{ab}}\,{\bf h}^{ab}(\varrho) & \ov\omega \end{array}\right],
\end{equation}
(we used the fact that $\ov\w=\w^T$) and ${\bf h}^{ab}(\varrho)$ denotes the $r\times r$ 
matrix whose elements are $h^{ab}_{\alpha\beta}(\varrho)$.



\section{Calculation for Werner states}\label{WernerStates}
In this Section we apply the developed statistical method 
to study Werner states of a $2 \otimes 2 $ dimensional 
system. They are defined as follows:
\begin{equation}
W(p):=(1-p)|\Psi_-\rangle\langle \Psi_-| 
+ \frac{p}{4} {\bf 1}_2 \otimes {\bf 1}_2, \label{WernerS} 
\end{equation}
where:
\begin{equation}\label{BellB}
\ket{\Psi_\pm}:=\frac{1}{\sqrt{2}}\big(\ket{01}\pm\ket{10}\big),\quad
\ket{\Phi_\pm}:=\frac{1}{\sqrt{2}}\big(\ket{00}\pm\ket{11}\big)
\end{equation}
are the Bell basis states and $\{\ket 0,\ket 1\}$ is the 
standard basis of $\C^2$.
The states $W(p)$ have positive partial transpose, and hence are 
separable (Peres and Horodecki \e \cite{PeresHoro}), for $p\ge 2/3$. 
As the fixed eigenensemble 
$\{\ket{e_{\alpha}}\}$ of $W(p)$ we take:
\begin{eqnarray}
& & \ket{e_1}:=\sqrt{1-\frac{3}{4}p}\,\ket{\Psi_-}, \quad 
\ket{e_2}:=\frac{\sqrt{p}}{2}\,\text{i}\ket{\Psi_+}, \quad \\
& & \ket{e_3}:=\frac{\sqrt{p}}{2}\,\text{i}\ket{\Phi_{-}},\quad 
\qquad \ \!\ket{e_4}:=\frac{\sqrt{p}}{2}\ket{\Phi_{+}}.
\end{eqnarray}

We proceed to calculate the one-particle partition
function $Z_1\big(\beta,\w;W(p)\big)\equiv Z_1(\beta,\w;p)$.
In what follows we assume $p>0$, for 
$p=0$ corresponds to a pure state. According to the general formula
(\ref{poHS}), we have to find the matrices
${\bf h}^{ab}\big(W(p)\big)$ and $M_{W(p)}(s,\w)$, defined in Eqs.~(\ref{hab}) and (\ref{macierz}). 
Since in the case of $\C^2\otimes \C^2$ the skew-symmetric subspace
$\C^2\wedge\C^2$ is one-dimensional---it is spanned by a single vector
$\ket\zeta=1/\sqrt 2 (\ket{01}-\ket{10})$ in each copy $AA'$ and $BB'$---there
is only one matrix ${\bf h}^{ab}\big(W(p)\big)\equiv {\bf h}(p)$ and only one 
Hubbard-Stratonovich parameter $s_{ab}\equiv s$.
Calculation of ${\bf h}(p)$ and $M_{W(p)}(s,\w)\equiv M_p(s,\w)$ yields:
\begin{eqnarray}
& & {\bf h}(p)=\frac{1}{8}\left[\begin{array}{cccc} 
4-3p & 0 & 0 & 0\\ 
0 & p & 0 & 0\\ 
0 & 0 & p & 0\\
0 & 0 & 0 & p \end{array} \right]\label{hp}\\
& & M_p(s,\w)=\left[\begin{array}{cc}
\omega & -2\text{i}s {\bf h}(p)\\
-2\text{i}\,\ov s \, {\bf h}(p) & \ov\omega \end{array}\right],\label{M}
\end{eqnarray}
so that:
\begin{equation}
E_1(\z;p)=\frac{1}{64}\big|(4-3p)z_1^2+pz_2^2+pz_3^2+pz_4^2\big|^2,
\label{F1}
\end{equation} 
and:
\begin{eqnarray}
Z_1(\beta,\w;p)=\int\de ^2z_1\dots \de ^2z_4
\text{exp}\Big[-{\beta}\,
\big|(4-3p)z_1^2+pz_2^2+pz_3^2+pz_4^2\big|^2 
-\langle \z|\w \z\rangle+\text{tr}\w\Big]\label{Z1werner}
\end{eqnarray}
(we have absorbed the factor $1/64$ into the definition of 
the parameter $\beta$). 

Next, we calculate $\text{det}M_p(s,\w)$ for $p \ne 0$. 
We first perform a transformation: 
\begin{eqnarray}
M_p \mapsto M_p':=
\left[\begin{array}{cc}
{\bf h}(p)^{-1/2} & 0\\ 
0 & {\bf h}(p)^{-1/2} \end{array}\right]\,
M_p
\left[\begin{array}{cc}
{\bf h}(p)^{-1/2} & 0\\ 
0 & {\bf h}(p)^{-1/2} \end{array}\right]
= \left[\begin{array}{cc}
\omega' & -2\text{i}s\\
-2\text{i}\, \ov s & \ov{\omega'}\end{array}\right],
\label{M'}  
\end{eqnarray}
where: 
\begin{equation}\label{w'}
\omega': = {\bf h}(p)^{-1/2}\omega {\bf h}(p)^{-1/2}\ .
\end{equation}
Then we multiply Eq.~(\ref{M'}) on the left by 
$\left[\begin{array}{cc}
{\bf 1} & 0\\ 
2\text{i}\,\ov s & \omega' \end{array}\right]$ 
to obtain:
\begin{equation}
\left[\begin{array}{cc}
{\bf 1} & 0\\ 
2\text{i}\,\ov s & \omega' \end{array}\right]
M_p' 
= \left[\begin{array}{cc}
\omega' & -2\text{i}s\\
0 & 4|s|^2+\omega'\,\ov{\omega'}\end{array}\right],
\end{equation} 
and after taking the determinants of both sides:
\begin{equation}
\text{det}M_p(s,\w) =\text{det}{\bf h}(p)^2\,\text{det}
\big(4|s|^2+\omega'\,\ov{\omega'}\big).\label{detM}
\end{equation}   
We then substitute Eq.~(\ref{detM}) into Eq.~(\ref{poHS}) and 
finally obtain (with $x:=4|s|^2$):
\begin{equation}
Z_1(\beta,\w;p)= \frac{\pi^4}{4\beta\,\text{det}{\bf h}(p)} 
\, \text{e}^{\text{tr}[\omega'{\bf h}(p)]} 
\int\limits_0^{\infty}\!\!\frac{\de x \,\text{e}^{-\frac{x}{4\beta}}}
{\sqrt{\text{det}\big(x+\omega'\,\ov{\omega'}\big)}},\label{Z1HS3}
\end{equation}
where $\w'$ is defined through Eq.~(\ref{w'}). The above integral is well defined, since 
$\text{det}\big(x+\omega'\,\ov{\omega'}\big)=
\text{det}\big(x+\sqrt{\omega'}\,\ov{\omega'}\sqrt{\omega'}\big)$ 
and $\sqrt{\omega'}\,\ov{\omega'}\sqrt{\omega'}$ is strictly positive, as
we have assumed that $\text{det}\w\ne 0$. We can explicitly calculate the 
derivative $\partial \text{log}Z_1 (\beta,\w;p) / \partial \w'$.
For a generic $\w'$ it takes the following form:
\begin{eqnarray}
& &\frac{\partial \text{log}Z_1 (\beta,\w;p)}{\partial \w'}=
\sqrt{{\bf h}(p)}\frac{\partial \text{log}Z_1 (\beta,\w;p)}{\partial \w}
\sqrt{{\bf h}(p)}\nonumber\\
& &=\Bigg[\int\limits_0^{\infty}\!\!\frac{\de y \,\text{e}^{-\frac{y}{4\beta}}}
{\sqrt{\text{det}\big(y+\omega'\,\ov{\omega'}\big)}}\Bigg]^{-1}
\int\limits_0^{\infty}\!\!\frac{\de x \,\text{e}^{-\frac{x}{4\beta}}}
{\sqrt{\text{det}\big(x+\omega'\,\ov{\omega'}\big)}}
\Big[{\bf h}(p)-\big(x+\omega'\,\ov{\omega'}\big)^{-1}\w'\Big]\label{dZ}.
\end{eqnarray} 
The special case of 
Eqs.~(\ref{Z1HS3}), (\ref{dZ}) for Bell-diagonal states is straightforward---it 
is enough to replace matrix ${\bf h}(p)$ from Eq.~(\ref{hp}) with the diagonal
matrix $4\,\text{diag}(1-p_1-p_2-p_3,p_1,p_2,p_3)$. 

\begin{figure}
\begin{center}
\includegraphics[width=0.55\linewidth, angle=270] {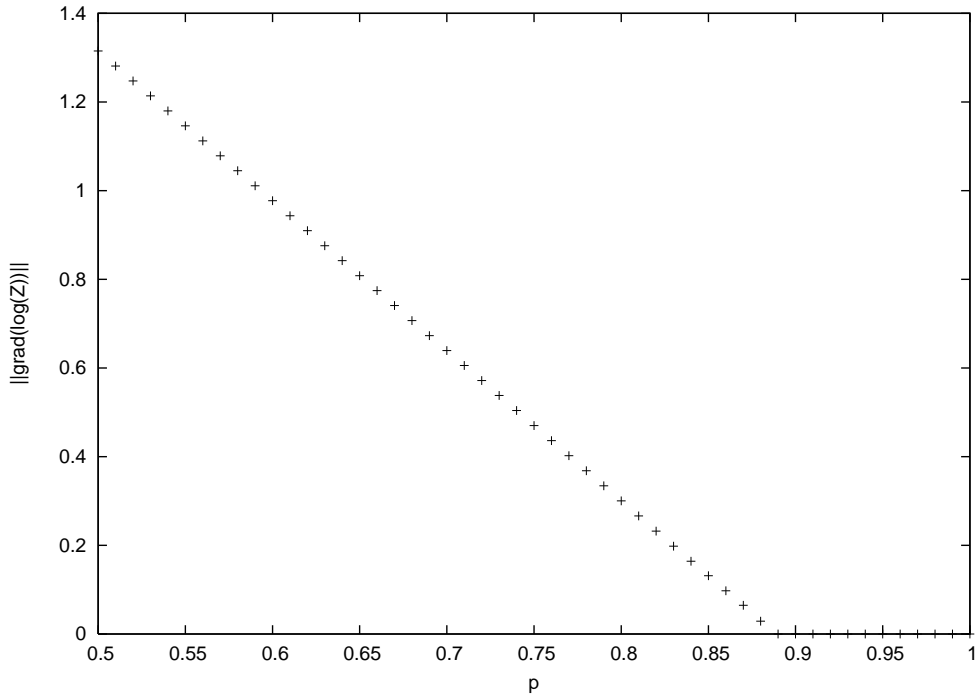}
\caption{\label{wykres1}The plot of 
$\min\limits_{\w'}||\partial \text{log}Z_1(\beta,\w;p)/\partial \w'||_{HS}$ 
for Werner states as a function of probability $p$ for $\beta=10$.}
\end{center}
\end{figure}

\section{Numerical results}
Further studies of the integral (\ref{Z1HS3}) were performed
using numerical methods. 
According to Eq.~(\ref{constr}) one has to search for a saddle 
point of $\text{log} Z_1(\beta,\w;p)$ with respect to $\omega$
(or equivalently with respect to $\omega'$; cf. Eq.~(\ref{w'})). 
The search was performed by flood-minimizing the Hilbert-Schmidt norm of 
$\partial \text{log}Z_1 (\beta,\w;p) / \partial \w'_{\alpha\beta}$
for a range of parameters $\beta=10,100,\dots$.
For simplicity we assumed a specific form of $\w'$: 
\begin{equation}\label{choice}
\w'=\left[\begin{array}{cccc} 
\gamma & 0 & 0 & 0\\ 
0 & \lambda & 0 & 0\\ 
0 & 0 & \lambda & 0\\
0 & 0 & 0 & \lambda \end{array} \right] 
\end{equation}
and minimized the derivative (given by formula similar to
to Eq. (\ref{dZ}), but taking into account the specific symmetry of (\ref{choice}))
with respect to the parameters $\gamma,\lambda >0$. We payed attention that 
the obtained minima are not on the border of the region $\omega' >0$
(or equivalently $\omega >0$). 
The specific choice (\ref{choice}) of $\w'$ was motivated by the form of the cost function
(\ref{F1}). We also obtained some numerical evidence that in the generic case
the minima of $||\partial \text{log}Z_1 (\beta,\w;p) / \partial \w'||_{HS}$
were attained for matrices $\w'$ very close to (\ref{choice}). 
The results of the simulations for $\beta=10$ are presented in Fig.~\ref{wykres1}
(the results for higher values of $\beta$ did not 
differ from those for $\beta=10$). We see that for 
$p\ge 0.89$ the constraints (\ref{constr}) can be satisfied. 
We shall call the interval where it happens ``equipartition region''.

\begin{figure}
\begin{center}
\includegraphics[width=0.55\linewidth, angle=270]{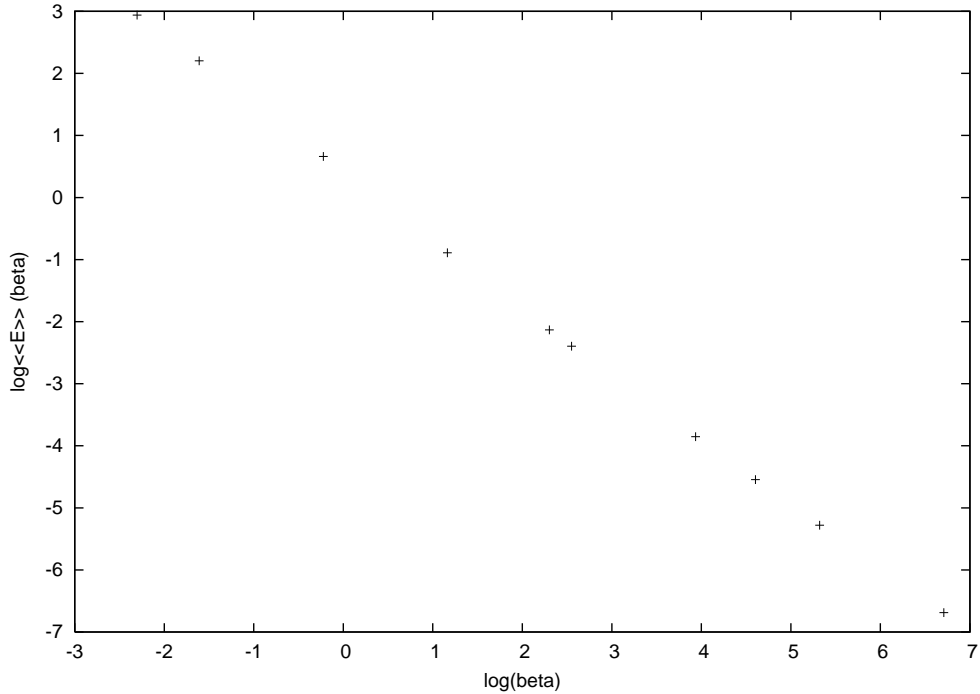}
\caption{\label{wykres2} The plot of 
$\langle\langle E_{1W(p)} \rangle\rangle_0(\beta)$ 
for $p=0.90$ on a double log scale.}
\end{center}
\end{figure}
Next, the dependence of the average entanglement 
$\langle\langle E_{1W(p)} \rangle\rangle_0(\beta)$ 
(cf. Eq.~(\ref{avH})) 
of the continuous ensemble (\ref{cont_ens}) on $\beta$ within the 
equipartition region was examined (recall that outside this region the  
one-particle constraints (\ref{cont_ens}) are no longer satisfied). Fig. 
\ref{wykres2} shows a sample plot for $p=0.9$. One sees that 
the average ``energy'' indeed scales like 
$1/\beta$, just like predicted by the Ansatz (\ref{ansatz})
and Eq.~(\ref{1/b}). The estimated exponent $\delta$ at this 
value of $p$ is $\delta\approx 1.75$.
We have also checked that in the limiting case 
$\beta \to \infty$ the equipartition region is not altered. Hence, 
our procedure seems to detect separability of the Werner states 
(\ref{WernerS}) at least for $p \ge 0.89$ 
and thus can serve only as a sufficient 
condition for separability.
We did not check the behavior of 
$\langle\langle E_{1W(p)} \rangle\rangle_0(\beta)$ outside 
the equipartition region $p<0.89$.

\section{Further questions and concluding remarks}
The statistical mechanical approach to the separability problem as
presented here differs from the more traditional techniques in that we studied the 
space of convex decompositions of a given state, rather than the 
convex set of all states. 
The resulting polynomial equations are real 
due to the constraint (\ref{wiaz}) and this real structure makes the 
analysis more complicated than it would be in a complex case. 
Hence, we applied statistical-mechanical methods to study 
possible zeros of this system. As an example we studied
$2\otimes 2$ Werner states (\ref{WernerS}). 
However, the numerical difficulty already at this simple example 
was quite high and we have applied several simplifications.
Nevertheless, the numerical results
suggest that 
at least for separable states in a vicinity of the identity, 
the partition function and the average ``energy'', related to 
the ensemble entanglement (cf. Eq.(\ref{p^2}),
show some qualitative change in their behavior. 

There are obviously some important questions left.
First of all, we postulated rather than derived 
the power-law state density behavior (\ref{1/b})
for entangled states.
It would be an interesting, albeit difficult, task
to try to analytically derive this law. Or at least to
find some arguments in its favor. 

Another thing is that in passing
from the full $N$-particle constrains (\ref{constr_stare}) to 
the one-particle one (\ref{constr}) we have tacitly assumed 
a sort of ``equipartition'' of the constraints, 
i.e. that the constraints are divided equally
among the particles. But it actually does not have to be like that.
In particular, the shape of the curve in Fig.~\ref{wykres1}
tells us that below $p=0.89$ the constrains are not ``equiparted''.
Thus, in principle one should work with the full $N$-particle 
partition function (\ref{Z_stare}) and seek regions were
full constraints (\ref{constr_stare}) can be satisfied.
Then the scaling of the average ``energy'' with $\beta$ within 
that regions will be able to discriminate between separability and entanglement.
  
As a side remark, we note that quite surprisingly, the value
$p=0.89$ appears in Braunstein \e~\cite{Braunstein} separability criterion,
based on an estimation of the size of a ball of separable states around 
the normalized identity 
(see also Bengtsson and $\dot{\text{Z}}$yczkowski~\cite{Bengtsson}
and the references therein). It will be worth analyzing this curious coincidence
in order to gain a deeper understanding of the strengths and weak points of the presented approach.

Finally, let us mention that in principle one can try to 
directly numerically calculate integral (\ref{Zinna}) using Monte Carlo method.
The points of $V_{N,r}$ can be generated either using Eq.~(\ref{Srozw})
or, what seems more feasible, directly from definition (\ref{st}). 
The latter method amounts to generating random unitary matrices 
from $U(N)$ and discarding $(N-r)$ of their columns (for the methods of
random generation of unitary ensembles see e.g. Po\'zniak \e~\cite{Pozniak}).
However, we have
not performed such simulations.

We gratefully acknowledge discussions with H.-U. Everts, 
P. Horodecki, G. Palacios, and R. Wimmer.
We would like to thank Deutsche Forschungsgemeinschaft
(SFB 407, SPP 1078, GK 282, 436 POL),
the European Graduate College 665, 
EU IP project SCALA, ESF PESC Program QUDEDIS, Spanish MEC Program Consolider Ingenio 2010, 
and Trup Cualitat Generalitat de Catalunya 
for the financial support.


\begin{thebibliography}{99}
\bibitem{GS} The Gram-Schmidt orthogonalization creates an orthonormal 
basis of a span of a set $\{\ket{v_1},\dots,\ket{v_r}\}$ of a linearly independent vectors
form some Hilbert space. Define:
\begin{equation}
\ket{\widetilde{u}_1}:= \ket{v_1},\quad 
\ket{\widetilde{u}_i}:=\ket{v_i}-\sum_{j=1}^{r-1}\frac{\langle
\widetilde{u}_j|v_i\rangle}{||\widetilde{u}_j||^2}
\,\ket{\widetilde{u}_j}.\nonumber
\end{equation}
Then 
$\{\widetilde u_1,\dots,\widetilde u_r\}$ is a orthogonal
system and spans the same space as $\{v_1,\dots,v_r\}$. Passing to the 
normalized vectors: 
$\ket{u_i}:=\frac{1}{||\widetilde{u}_i||^2}\,\ket{\widetilde{u}_i}$,
we obtain the desired orthonormal system.
\end{thebibliography}
\end{document}